\documentclass[sigconf,nonacm]{acmart}
\usepackage[utf8]{inputenc}
\AtBeginDocument{%
  \providecommand\BibTeX{{%
    \normalfont B\kern-0.5em{\scshape i\kern-0.25em b}\kern-0.8em\TeX}}}

\copyrightyear{2020}
\acmYear{2020}
\acmDOI{}
\acmConference{arXiv}{July 16}{2020}
\acmPrice{}
\acmISBN{}
\thanks{LA-UR-20-23043}
\thanks{The U.S. Government retains for itself, and others acting on its behalf, a paid-up nonexclusive, irrevocable worldwide license in said article to reproduce, prepare derivative works, distribute copies to the public, and perform publicly and display publicly, by or on behalf of the Government. The Department of Energy will provide public access to these results of federally sponsored research in accordance with the DOE Public Access Plan. http://energy.gov/downloads/doe-public-access-plan}

\title{Modernizing the HPC System Software Stack} 

\author{Benjamin S.~Allen}
\affiliation{%
    \institution{Argonne National Laboratory}
    \city{Lemont}
    \state{Illinois}
}

\author{Matthew A.~Ezell}
\author{Paul Peltz}
\affiliation{%
    \institution{Oak Ridge National Laboratory}
    \city{Oak Ridge}
    \state{Tennessee}
}

\author{Doug Jacobsen}
\author{Eric Roman}
\affiliation{%
    \institution{Lawrence Berkeley National Laboratory}
    \city{Berkeley}
    \state{California}
}

\author{Cory Lueninghoener}
\author{J.~Lowell Wofford}
\affiliation{%
    \institution{Los Alamos National Laboratory}
    \city{Los Alamos}
    \state{New Mexico}
}

\date{\today}

\begin{document}

\begin{abstract}
Through the 1990s, HPC centers at national laboratories, universities, and other large sites designed distributed system architectures and software stacks that enabled extreme-scale computing.  By the 2010s, these centers were eclipsed by the scale of web-scale and cloud computing architectures, and today even upcoming exascale HPC systems are magnitudes of scale smaller than those of datacenters employed by large web companies.  Meanwhile, the HPC community has allowed system software designs to stagnate, relying on incremental changes to tried-and-true designs to move between generations of systems.  We contend that a modern system software stack that focuses on manageability, scalability, security, and modern methods will benefit the entire HPC community. In this paper, we break down the logical parts of a typical HPC system software stack, look at more modern ways to meet their needs, and make recommendations of future work that would help the community move in that direction.

\end{abstract}


\maketitle


\section{Introduction}
\label{section:introduction}

The Department of Energy's (DOE) National Laboratories have long histories with high-performance computing, housing some of the largest HPC systems in the world. In the early to mid-1990s, the laboratories and HPC facilities defined large scale computing and created the tools and concepts that enabled the extreme-scale computing that we see in companies like Amazon, Facebook, and Google.

By around 2010, HPC systems were starting to be eclipsed by these companies, and today even upcoming exascale systems are magnitudes of scale smaller than those of datacenters employed by large web companies.  Meanwhile, the HPC community has allowed system software designs to stagnate, relying on small incremental changes to tried-and-true designs to move us between generations of systems.  While this can partly be attributed to the differences between the systems in question---tightly-coupled HPC systems versus widely distributed web systems, for example---we contend that a more modern system software stack that focuses on manageability, serviceability, scalability, resiliency, security, and modern methods will benefit the entire HPC community.

Modernizing the system software stack used on HPC systems is not a simple task, and we will not try to solve it in this short paper.  While existing HPC system management packages like xCAT\cite{xcat} and Warewulf\cite{warewulf} have made incremental steps towards the goals described in this paper and new efforts like Shasta\cite{shasta} have made some of them explicit design goals, this is long-term work that must be done as a community.  Designing and implementing a full modern HPC system software stack will take multiple years to get right, and if treated as a monolithic project will be obsolete by the time it is complete.  Instead, we are starting the work of breaking the problem down to create a foundational design, and to iterate over this design's subcomponents.

In this paper, we will look at the base components in today's system software stack, describe the improvements we see each component needing to work in a more modern way, and survey the current state of work on those components.  We have narrowed our discussion down to two broad areas of interest to teams that manage HPC systems: the software stack that runs on individual cluster nodes, and tools and practices involved in system management activities.

While each of these areas is unique, they can be discussed using a similar set of topics.  These include manageability and serviceability, scalability and resiliency, and security.  In addition to these topics, we discuss how API-driven designs and applying modern methods can help appropriate areas, and look at the current state and potential implementations in the more concrete areas.

\subsection{Paper Layout}
\label{subsection:overall-model}

This paper consists of two main sections.  In Section \ref{section:node-stack}, we describe components of compute and service nodes within an HPC system.  This is comprised of sections discussing using a minimal operating system stack as the basis for compute and service nodes; ways of running cluster support services on top of that minimal operating system; and ways of running compute jobs on top of that minimal operating system.  In Section \ref{section:system-management}, we look at higher-level activities that occur on an HPC system and how they can benefit from updated approaches.  This is comprised of sections discussing configuration management, state management, orchestration, and provisioning.

\section{Node Stack}
\label{section:node-stack}

Nodes contained within a cluster can largely be categorized under two types: {\it compute nodes}, which run user jobs, and {\it service nodes}, which provide system management, user access, data routing, and other services to a system.  The software stack used on either type of node should focus on reusable components, the rapid boot of nodes, and flexibility for both service and user job containerization.

This section describes components within the operating system and software stack that runs on top of these nodes.  Section \ref{subsection:minimal-os} starts with a description of the common minimal operating system architecture that will be used as a basis for either compute or service nodes.  Section \ref{subsection:services} then describes how system services run on the cluster, building on the minimal operating system framework.  Finally, Section \ref{subsection:jobs} describes compute resources and how jobs will be executed on those systems.

\subsection{Minimal OS}
\label{subsection:minimal-os}

Individual compute nodes in current HPC systems are generally stateless, whether they include persistent storage that can be rebuilt at any time (e.g.~state-lite) or because they lack persistent storage and lose all state when powered off.  True stateless systems typically boot with a staged approach: in stage one, a kernel and initial RAM disk ({\it initrd}) are loaded via a network device, while in stage two a permanent root filesystem is either downloaded or mounted over the network and the kernel pivots its root to that new image.

We advocate using a minimal operating system image that does not abandon the initial RAM disk image via the switch root mechanism.  Instead, the initrd image should include the minimal set of binaries, libraries, and services that are needed to support building more advanced environments using containers.  These containers should make use of existing foundational technologies like Linux namespaces\cite{namespaces} and container runtimes built upon these technologies.

\subsubsection{Current State} 

Past and current projects have implemented some of the approaches described in this section.  Lightweight operating system kernels such as the Blue Gene Compute Node Kernel\cite{cnk} and Catamount\cite{catamount} have been used in the past, but these are specialized kernels that are meant to run a single process very quickly and are not full-featured operating systems.  Several minimal Linux distributions have been developed to power container-based web applications, such as CoreOS Container Linux\cite{coreos}, RancherOS\cite{rancheros}, and openSUSE MicroOS\cite{microos}, but these distributions are generally targeted toward microservices environments and don't take into account needs of HPC systems, such as high-speed network adapter drivers and parallel filesystems.

\subsubsection{Manageability \& Serviceability}

By using a minimal base operating system on cluster nodes, several advantages will be gained over heavier-weight OS images.

\begin{description}
  \item[Reduced code base] In a monolithic stateless system, updating nearly any software component requires rebooting into a new image to ensure a fully updated system.  By moving to a minimal OS image with containerized services and user jobs, we can dramatically reduce the number of changes that require reboots to include only the kernel and base services.  More discussion on how approach benefits jobs and higher-level cluster services can be found in Sections \ref{subsection:services} and \ref{subsection:jobs}.

  \item[Reduced image configuration] A reduced set of base packages and services reduces the amount of post-boot configuration that needs to be run in the OS image itself.  This will result in simplified node image configurations and lower node boot time while moving configuration logically closer to the containers of the applications being configured.

\end{description}

\subsubsection{Scalability \& Resiliency}

The minimal OS approach benefits scalability and resiliency by reducing the size of the initial system boot artifacts and simplifying the number of components that are active in those images.  While application and service container images still need to be transferred to nodes after their initial boot, this can be implemented with more scalable and more robust clients within the minimal image than within the traditional PXE or firmware framework.  Once the system is booted, a minimal OS design makes it easier to logically separate a node's base image from its applications and services, providing a layer for a center to introduce sandboxing and automatic remediation tools that are less likely to be affected by the applications themselves.

\subsubsection{Implementation}

The following components should be a part of any successful Minimal OS implementation.

\begin{description}
  
  \item[Kernel, Kernel Modules, and Hardware Support] The minimal OS should be based on a standard Linux kernel.  This provides flexibility and reliability to the underlying system and provides compatibility with a very large software ecosystem and developer base.  While much of the system functionality will be added by components or containers loaded outside of the minimal OS, the minimal OS is responsible for managing kernel-level features such as hardware driver modules, firmware loading, and filesystem drivers.
  
  \item[Initial ramdisk] Network booting techniques currently require an initial ramdisk as the first stage of OS loading to provide enough system functionality to complete the boot process.  For the minimal OS, the initrd should be comprised of a userland environment based on a lightweight toolset such as u-root\cite{u-root}.  The initrd must also be capable of acquiring and loading any service or job containers that need to run on top of it.
  
  \item[Read-only root filesystem image] The root filesystem image of the minimal OS, as provided by the provisioning system (see section \ref{subsection:provisioning}) should be provided in a way that does not allow persistent changes.  This helps to ensure that the system image remains stateless.  The read-only root filesystem can be overlayed with a read-write memory-based overlay filesystem\cite{overlayfs} to provide runtime functionality.
  
  
  \item[Boot-time OS configuation] Individual nodes within the system will need boot-time specialization, such as network configuration and hostname assignment.  These configuration items should be simple, but extensible, and able to be passed to the node via kernel command-line parameters, fetched remotely during boot, or provided via tools like cloud-init\cite{cloud-init}.
  
  
  
  
  
\end{description}

\subsection{Cluster Services}
\label{subsection:services}

On a traditional HPC cluster, there are one or more nodes that run internal services that manage the cluster.  While these nodes may be referred to by a variety of names (``master nodes'', ``leader nodes'', ``service nodes'', ``head nodes'', ``admin nodes''), in this paper we will refer to them generically as ``service nodes''.  Services that these nodes run include:

\begin{itemize}
  \item System support (e.g.~central NTP, Syslog, LDAP, authentication, mail)
  \item Job schedulers and resource managers (e.g.~Slurm, TORQUE, Moab, PBS, SGE, Cobalt)
  \item System provisioners and support (e.g.~Warewulf, Cobbler, DHCP, TFTP, HTTPD, NFS)
\end{itemize}

Some of these services, such as NTP and DHCP, are very simple and essentially stateless within an HPC cluster environment, while others, such as Slurm and LDAP may have running or on-disk state that needs to be preserved during runtime or across service restarts.  In all cases, these services normally run on one or more {\it service nodes}, which are independent of the cluster's compute nodes.  On a small cluster, all of these services may run on a single node, while on a larger cluster they may be spread out to several nodes that hierarchically manage subsets of the compute nodes.

While this design is simple to set up and manage, it comes with drawbacks.  For example, it creates single points of failure, in which a single node failure can cause subsets of the cluster or even the entire cluster to fail; it creates scaling constraints, where adding new hardware may require rebalancing the existing infrastructure or growing it by quantum leaps; and it adds artificial software upgrade constraints, in which large parts of a cluster may need to be taken out of service to perform upgrades on individual components.  In short, this design is easy but constraining.

To update this design, we can exploit the fact that many of these services are built to scale out by adding more independent copies of the service on multiple servers.  While this is essentially the path that is taken on larger clusters that have multiple service nodes, it is usually only used in this case to provide scaling without extra resiliency.  By using service management tools that were originally adopted by large-scale, microservice-oriented web companies, we can start to build an HPC cluster services design that is both scalable and more resilient than on a traditional cluster.  Along the path of adopting those tools, we can start to see how adopting stronger introspection tools also help make systems more robust.  Two key design decisions behind this plan are to make services independent and replicable, and to add visibility into the APIs they provide.

\subsubsection{Current State}

The services that run clusters are developed by many different people, projects, and companies, and their design and current state are highly varied.  Some, such as DHCP, DNS, and NTP servers, have long histories and are used across the computing industry. These tend to scale well because they were written to be run replicated multiple times at a site. Meanwhile, HPC centric services like Slurm, PBS, Conman, and Powerman were written to have one copy running on a ``master'' node without significant consideration on how to make multiple copies run on the same system at the same time.

Complex HPC services like schedulers, resource managers, and cluster management tools have frequently been built to scale vertically instead of horizontally.  Most of these tools could benefit from increased introspection and better resiliency, as they frequently become single points of failure within their clusters.  

While it is currently possible to build a fairly scalable and resilient system using off-the-shelf services and some high availability tools, it is not easy to get actionable monitoring data that can answer more complex questions than ``is it working'' off of many of the components.  Data on client and API call latency, time spent processing versus handling requests, and the number of requests served over time will provide useful monitoring and introspection data for system managers.

\subsubsection{Manageability and Serviceability}

Large-scale cluster manageability and serviceability are strongly influenced by the design of the cluster's services infrastructure.  A modern system needs better componentization of its services and more scalable introspection tools.  We can start to see how to do this by applying some standard concepts.

\begin{description}
  \item[Containerization/Virtualization] Containerization and virtualization give system managers flexible ways to upgrade service node images, migrate running services between service nodes, and scale services up or down as demand or system design changes.  Many low-level (e.g.~Docker and Libvirt) and high-level (e.g.~Openstack and Kubernetes) tools exist that can be used to provide this level of support.  By exploiting these tools, it is possible to build a service infrastructure on top of lightweight virtual machines or containers that can be treated as independent and ephemeral.
  \item[Minimal OS] The service infrastructure should make use of the same minimal OS described in Section \ref{subsection:minimal-os} as its underlying support operating system.  Having a consistent base operating system across the full system simplifies many aspects of system management, including node image building, security monitoring and patching, and general system debuggability and discoverability.
  \item[Service profiling] Services should provide means for profiling their execution.  For each type of request, a request-response style service should provide means for determining: the number of requests made, the total number of responses, the average service time per request, and the average response per request. In many applications other statistics (e.g.~median response time, 99\% response times and other distributional properties) are desirable.
  \item[Visibility into operations] Services should be capable of logging and reporting their activity during initialization, normal operations, termination, and other major state changes during execution.  Per-request operation logging should be provided for debugging purposes if possible.  Failed requests should be reported.  Typically debugging traces implemented for development purposes are also suitable for problem diagnosis for systems administration, and systems administrators should have the option to enable such logging messages.
\end{description}
  
\subsubsection{Scalability and Resiliency}

While smaller systems may not require scalability features that are necessary to run larger-scale clusters, systems of all sizes benefit from increased resiliency.  An important aspect of scaling services is improving their resiliency by making them more tolerant of failure and less dependent on other cluster services.

\begin{description}
  \item[Resiliency] Treating services as smaller, independent portions of a system introduces new levels of resiliency.  By using lightweight containers or virtual machines as the basis for each individual container and by aiming for stateless, ephemeral services when possible, we enable the ability to treat services as components that can be quickly started, restarted, and replaced without affecting a running system.  This also makes it less likely that a single service running on a larger system can affect other services on that system if it starts experiencing problems.
  \item[Failure Modes] A failure in a service node should not result in failures of client nodes.  Clients accessing services should be tolerant of failures of service nodes and either retry operations until a request is successfully processed or fail.  A client should be notified quickly if a server is unable to handle requests so that it can quickly send service requests to another server instance; active notification of server failures should be preferred to detecting faults via timeouts.
  \item[Transparent load balancing] Services that scale up and down automatically must not require a configuration management action on all client nodes to be usable by the client nodes.
  \item[Automatic scalability] Extra copies of services should be able to spin up and down automatically.  Small, independent services running within an orchestrated environment helps facilitate this.
  \item[Cluster independence] Ideally, the infrastructure described in this subsection could be used for more than one logical cluster at a time.  This would involve leveraging the scalability of the service layer to handle multiple independent (or pseudo-independent) clusters that have boundaries defined around processor architecture, high-speed network interconnection, or similar hard edges.
\end{description}

\subsection{Jobs}
\label{subsection:jobs}

Ultimately, an HPC system exists to support the running of user jobs across many systems at once.  These jobs are frequently scientific in nature, but they also support financial market analysis, oil and gas exploration, cinematic movie rendering, and many other efforts.

Today, running a complex job on an HPC system requires knowledge of what system-specific libraries, scientific libraries, and maybe kernel libraries are available on a particular system, as well as what their paths are and how they are linked with each other.  This software stack tends to be specific to individual systems, and is generally difficult to change at a system level due to deep dependency chains that reach into vendor-provided software and due to dependencies across many users' projects.  The community often works around this by using tools like SPACK\cite{spack} to build software dependencies in a project-local area, but this is far from ideal.

Containerization, referring to the use of a combination of Linux namespaces and filesystem trees to provide flexible software stacks on Linux systems, is a popular way to work around some of these problems in today's web industry.

\subsubsection{Current State}

Outside of the HPC world, containers based on Linux namespaces are widely used for rapid prototyping and deployment of stateless services.  Meanwhile, the HPC community has been slow to widely adopt the same technologies.

On one end of the implementation spectrum, full-service containerized scheduling and orchestration solutions like Kubernetes\cite{kubernetes} and OpenShift\cite{openshift} can be used to easily deploy and scale stateless services on large clusters of systems.

On the other end of the spectrum, very lightweight containerization tools like Charliecloud\cite{charliecloud}, NVIDIA's enroot\cite{enroot}, {\tt systemd-nspawn}, and the Linux {\tt unshare} tool provide simple containerization solutions.

Several tools span the gap between these two extremes, including ones that target HPC environments (such as Shifter\cite{shifter} and Singularity\cite{singularity}) and ones that target more generic workloads (such as Docker\cite{docker}).  These tools provide varying levels of support for scheduling, orchestration, and container storage.

\subsubsection{Usability}

Since we do not expect all users will need (or want) to become containerization experts, we propose that this functionality be provided in two ways:

\begin{itemize}
 \item First, allow significantly advanced users to bring their containers to run on a system.  Ideally, we should standardize on a single container image format that can either be used as-is or translated on-the-fly to work on any system.  Given the inability of a center to know the provenance of a given user-provided container, the processes in these containers must run unprivileged.

 \item Second, provide transparent containerization to anybody who does not provide their application container.  Many users will still desire a workflow where they log in to a cluster frontend node, compile their code with the provided compilers and libraries, and submit their job to the backend nodes.  In this case, the scheduler and resource manager should work with the cluster management system to transparently stand up a default center-provided container that matches the frontend node environment, and the job should run inside that environment without additional user intervention.
\end{itemize}

\subsubsection{Manageability \& Serviceability}

By using containers as an abstraction that separates a system's user environment from its underlying kernel, we realize several benefits that improve system manageability and serviceability.  These include:

\begin{description}
  \item[User Environment Upgrades] On traditional clusters, upgrading the system-level packages that make up the user environment requires some sort of downtime to either upgrade packages in-place or to distribute and boot new images that have been built out-of-band.  In a containerized environment, the process of distributing new images is dramatically simplified: once a particular job finishes, the next job can start with the new container image that has been independently distributed to the cluster.
  \item[User Environment Flexibility] For the same reason that upgrading operating system images can be difficult on a traditional cluster, providing multiple versions of a user environment is also difficult.  Using a transparently containerized model, the entire user environment can be chosen at job submission time, and multiple versions of the user environment can be active on the system at a given time.
  \item[Operating System Separation] By disconnecting the user environment and libraries from the operating system kernel itself, system managers and system architects can change the underlying operating system kernel with more freedom.  While current HPC clusters are frequently built using Redhat- or CentOS-based images for compatibility reasons, a fully containerized cluster can provide that compatibility layer within the user environment container while making use of the best base operating system for a given system.
\end{description}

\subsubsection{Implementation}

Any implementation of a transparently containerized workflow requires support from the major HPC schedulers and resource managers.  To provide a very basic level of support, this requires the ability to start jobs on compute nodes with a minimal set of Linux namespaces in use.  In this most basic model, a user that wants to run a traditional job submission will submit their job and rely on the resource manager to start their processes in an appropriate center-provided container image with at least the {\tt user} and {\tt mnt} namespaces.  Meanwhile, a user with an already-packaged application container will provide a container in their job submission, and the resource manager will start their job with that container image within the appropriate namespaces.

A more advanced implementation will include support for a container store.  In this case, user-provided and center-provided container images will be stored in a container registry, and the resource manager will have the ability to obtain a specified container image from this registry at job startup.  The user's interface for running a traditional-style job will not change in this implementation. However, a user providing their image will be able to upload the image to the registry and be able to specify the image name when submitting their job.

The Linux namespaces used within an implementation will depend on local policy.  For node-exclusive scheduling, a minimal set of namespaces can be used: {\tt mnt} and {\tt user}.  For shared-node scheduling, centers will likely want to provide process privacy with the {\tt pid} namespace and resource constraints with {\tt cgroups}.  More advanced use cases may also require additional isolation, such as network separation with the {\tt net} namespace.

While providing abstraction between the user environment and the underlying kernel sounds great in theory, HPC systems often have dependencies that cross that boundary.  Examples include GPU libraries and MPI implementations that depend on the underlying high-speed network.  Today, these problems are frequently worked around by injecting the system's libraries into a container at run time.  True container transparency will require much better support for dynamically making use of tools and libraries that cross this abstraction boundary.

\section{System Management}
\label{section:system-management}

The HPC system described thus far requires the coordination of many components to deliver a usable system for running jobs.  The hardware must be provisioned, the containers must be made available, and basic operations such as software updates, access control changes, or system maintenance must be made straight-forward to perform while ensuring that operations do not cause damage to running workloads, to hardware, or other components of the system.  Systems management refers to the components that make these kinds of operations available.  We contend that the growing complexity and scale of systems, both in software and hardware components, demands a better design for HPC systems management.  Further, we observe that many of these systems management tasks can be informed by large-scale web and cloud techniques.

In this section, we outline the logical components for future HPC systems management.  We break down the logical components as: configuration management, state management, orchestration, and provisioning.  We stress that this is a logical separation of tasks rather than a reference to any specific pieces of hardware.  In some cases, it may be, that multiple or all of these components are provided by a single piece of software.  Nevertheless, we consider these distinct roles that must be considered in the design of future HPC systems management to achieve the stated objectives of manageability, scalability, and security.

Section \ref{subsection:configuration-management} begins by describing how adopting new approaches surrounding the management of configuration states can help modernize system management practices. Section \ref{subsection:state-management} then proposes methods for managing current and intended state in a distributed HPC system. Section \ref{subsection:orchestration} ties both of those together by discussing methods for translating a system's current state into its intended state through orchestration. Finally, Section \ref{subsection:provisioning} looks at the tooling and methods required to automate, orchestrate, and secure the system provisioning process leveraging the methods outlined in the previous sections.

\subsection{Configuration Management}
\label{subsection:configuration-management}
Using automated tools to configure systems has a long history\cite{cfengine} and is established as a best practice\cite{gettingstartedwithcm}. Configuration management systems are especially suited to managing HPC systems, which generally have a large number of very similar constituent nodes. Today, configuration management tools such as Ansible\cite{ansible}, Cfengine\cite{cfengine}, and Puppet\cite{puppet} are widely used in HPC environments, but their installations can benefit from an updated look at their roles and responsibilities.

\subsubsection{Manageability \& Serviceability}

Configuration management tools are designed to enhance the manageability and serviceability of a system by providing an automated method of making changes and enforcing policy. One strong case for using a configuration management tool is disaster recovery: in the case of a minor or major system disaster, a site needs to be able to return the system to the state it was in before the disaster. Flat-file configuration data facilitates this by being easy to back up, transfer, and repair, but not all services within an HPC system use flat files for configuration and state management. In this case, systems should provide an API or other interface that can be used by an arbitrary configuration management system to bootstrap the service with data stored in a more accessible format.

Given the ubiquity of configuration management tools within HPC centers, HPC systems and system software need to support a layered approach to system configuration management. When standing up a new system, it is undesirable to reimplement an existing configuration management scheme to comply with the new system's prescribed method of configuring software. While vendors and tools need to provide their base configurations, sites also require access to APIs or other interfaces that can be used to override or supplement these base configurations with site-specific configuration without having more than one competing source of truth.

Similarly, sites that make use of test and development environments alongside production environments should be able to specify site-wide configuration data that is inherited by all systems. This prevents the need to replicate that information between multiple systems, reducing errors. The deltas between production and development systems should be made relatively small and should be easily described and templated.

\subsubsection{Scalability \& Resiliency}

HPC systems with more than 20,000 nodes are typical in leadership class facilities, and configuration management tools need to scale to that size and beyond. Configuration operations must be asynchronous to one another so that there are no blocking operations while waiting for lagging portions of the system to catch up. Configuration systems must also be robust to failure of individual systems, capable of bringing a node that has been down for an extended period back to the same level of configuration shared by the rest of the system. Configuration should happen as quickly as the system will allow it to, adhering to individual tools' deployment models and design principles as closely as possible for consistency.

\subsubsection{Modern Methods}

Following current best practices\cite{infraascode}, system configuration should be definable by text files that can be managed by a source code management or version control system such as Git. Following this paradigm enables large operations teams to develop new changes on a system without interfering with the running system or other team members' work.

Following this paradigm enables several positive outcomes, including:

\begin{description}
  \item[Code review] By treating configuration data like software source code, we can use code review tools available with many revision control systems to review changes before they are applied to a system.
  \item[Safer team collaboration] Keeping configuration in a revision control system prevents many interference problems that can arise when working on large teams. The revision control system can be treated as the source of truth for the system's configuration, providing a clear interface for the system management team to use when making changes to the system.
  \item[Audit trails] Revision control systems provide a history of all changes made to a repository, providing an audit trail for security investigations, pinpointing a bad change, or other similar activities. A well-maintained repository can also simplify the process of reverting a change that had undesirable consequences, even if the change was made long in the past. Version control systems also allow for tracing the history of changes through the lifetime of the system.
\end{description}

\subsubsection{Security}

An important subset of data managed by a configuration management tool is security-related configuration. The proper configuration of these items, including SSH keys, authorization and authentication configuration files, and host-based firewall rules, is enforced by a configuration management tool. These tools also prevent unmanaged changes, whether malicious, accidental, or intentional, from persisting on a system. To reap the security benefits of a configuration management tool, the tool must be able to fully enforce the state of the system. Any state that cannot be defined by a configuration management tool can potentially be tampered with in a way that is not easy to automatically detect and remediate.

A configuration management tool used on a modern HPC system must include integrated secrets management. Whereas earlier generations systems were protected by relatively simple reusable or one-time password-based schemes to control logins and privilege escalation, newer systems make use of role-based access models, API keys, x509 certificates, and similar tools and technology to regulate authentication and authorization. The numerous keys, certificates, and tokens used on a system must be tracked, protected, and revoked and regenerated as necessary. A configuration management tool used in such a system must be able to encrypt secrets at rest and distribute secrets to systems without adding undue complexity. These secrets must be overridable by a site as needed to comply with their local security policies.

\subsubsection{Current State}

Configuration management is a mature field with many well-established products.  Most current configuration management tools assume they will be working with flat files or templates, and they generally do not have any problem with configuration repositories that are stored in a revision control system.  Hierarchical configuration control is available in tools like Puppet's Heira\cite{puppet} and the Ansible\cite{ansible} group structure, providing layered configuration management support.

Interoperability between existing configuration management tools at sites and those provided by vendors continues to be difficult.  Sites with systems from multiple vendors face the task of integrating multiple potentially conflicting configuration management tools with their site-deployed tools, which can lead to islands of one-off systems that are managed differently from the others at the site.  Secrets management also continues to be difficult.  While simple tools like Ansible Vault\cite{ansible} and HashiCorp Vault\cite{hashicorpvault} are sufficient for handling secrets, vendor adoption has been incomplete and integrating site-specific secrets management tools with vendor-provided configuration management tools is often not straightforward.

\subsection{State Management}
\label{subsection:state-management}

The successful design of any large, distributed system requires a coherent and reliable way to determine both the current and intended state of critical system hardware and software components.  In traditional HPC solutions, the states of particular components have been tracked by independent services and processes that do not cooperate within a shared state namespace.  This makes reliable automation and complex system workflows cumbersome or incoherent.  For instance, to achieve a basic function such as a rolling system update, one needs reliable answers to questions like ``is this node ready for reboot?''.  Without a coherent and reliable state store, these kinds of questions cannot be adequately answered to achieve reliable system management.  We contend that proper state management is critical to future HPC system management stacks.

As HPC systems grow in scale and complexity---both hardware and software---it becomes critical to system management to provide reliable automation and coordination of system tasks.  Even the common task of provisioning a node has become a complex workflow in today's multi-architecture and multi-imaged systems.  To achieve coherent, complex workflows both the current state of the system and the desired state of the system must be reliably known.  This maps closely to the declarative concepts of ``facts'' and ``desires'' (or ``goals'').  In a distributed system, the facts and desires need to be coordinated to formulate coherent answers to any question needed to achieve the desired workflow.

In general, the desired states of the system should be derived from, but are not necessarily identical to the configuration management system (see \ref{subsection:configuration-management}).  For instance, the configuration management system may provide the recipes for a minimal OS image, and specify the current approved image, which the state management system will register as a particular desire for a particular node.  The source of truth for system facts should ultimately be the hardware or software components themselves, or their nearest controlling process.  For instance, the best source of truth of the current power state of a particular node is either the node itself or its out-of-band controller.  For software and system management states, it is expected that this information may be provided by the orchestration components (see \ref{subsection:orchestration}), or the scheduler and resource manager (``SRM'').  In less complex systems, state management may be a provided component of orchestration or SRM.

\subsubsection{Manageability \& Serviceability}

By providing a coherent approach to state management, the system management architecture can provide known guarantees to the state of the system.  This allows not only the reliable and safe operation on the system by administrators of the system but also allows for the creation of complex orchestration and automation workflows.  The ability to create safe automation and orchestration of components will provide critical services to the rest of the system management stack, and is considered critical for the management of large--scale future systems.

\subsubsection{Scalability \& Resiliency}

When managing distributed state, consistency demands and guarantees must be considered.  For instance, if it is determined that absolute consistency is required for a particular question, the general guidelines outlined by the CAP theorem\cite{captheorem} must be weighed.  In these cases, a two-phase commit process, such as those provided by Paxos\cite{paxos} and Raft\cite{raft} algorithms must be followed.  These algorithms have been widely deployed in cloud and web-scale environments, and are freely available in packages like ZooKeeper\cite{zookeeper}, etcd\cite{etcd}, and Consul\cite{consul}.

While these algorithms can provide strong consistency guarantees, they tend to scale poorly to very large scale and high-transaction deployments\cite{ailijiang2016consensus}.  Fortunately, cloud and web-scale deployments have shown that, for a large class of system management concerns, it is acceptable to rely on ``eventual consistency,''\cite{10.1145/1435417.1435432} in which consistency is not guaranteed, but guarantees on consistency convergence over time can be provided.  For instance, in a large scale provision of a system, it may be acceptable that the wrong boot configuration is temporarily provided to a node so long as the error is caught and corrected within an acceptable timeframe.  The eventual consistency paradigm has shown strong scalability advantages for large scale distributed databases and systems management, making eventual consistency desirable for large-scale HPC management when strong consistency is not required. 

\subsubsection{Implementation}

Successful state management relies on the following requirements:

\begin{description}
  \item[Source of truth] All state information must have a consistent and well-defined source of truth.  If at any point the source of truth for a particular fact shifts, the source of truth for that fact must be reliably transferred, e.g.~the state of a node may initially be owned by the node's provisioner, then handed off to the SRM for job scheduling.  In general, the source of desires should be configuration, while the source of facts is components.
  \item[Defined consistency] For automation or orchestration to work effectively, all facts must have consistency guarantees or consistency convergence appropriate for the query.  For instance, the query ``can this node be powered off?'' may need a strongly consistent answer to avoid lost information, while the query ``which image should this node run?'' may have less strict consistency requirements so long as orchestration can converge the node to the correct image within acceptable timeframes.
  \item[Centralized] To provide consistent answers to queries requiring multiple facts, a centralized interface for state information must be provided.  In the case that multiple state management systems exist, they must manage disjoint sets of state, and no query across disjoint sets of facts should be considered coherent.
  \item[Exclusivity] Any process that access or supplies state must interact with the state manager for that information.
\end{description}

\subsubsection{Current State}

While state management solutions for cloud and web-scale solutions are common practice and readily available, state management has largely been ignored by current HPC system management solutions.  The one exception to this rule is the role the SRM currently plays in state management.  In some less-complex deployments, the SRM, along with some basic ties into orchestration, may be sufficient to supply HPC state management.  However, in most large-scale or complex deployments, HPC tooling will need to be adapted to use these techniques.

Many software packages, some already mentioned, provide adequate data stores with varying levels of consistency guarantees for the use with state management.  Whether one of these solutions is used, or an integrated solution is adopted, the orchestration, provisioning, and configuration management tooling will need to be adapted to consistently use a state management solution.  Some attempts at this already exist, such as Kraken\cite{kraken}. 

\subsection{Orchestration}
\label{subsection:orchestration}

Orchestration is the logical follow-on to state management. While state management tracks current and desired states of the system, the role of orchestration is to coordinate the changes of states in the system.  Orchestration is intended to provide coordinated, safe operations on the system.  For instance, orchestration should be responsible for coordinating tasks like system initialization, and rolling software updates.  Orchestration is responsible for the logic to make sure these tasks are performed in safe, robust, and coherent ways.

We consider orchestration distinct from the actual tooling to make system changes, for example, booting nodes.  Orchestration provides the logic necessary to coordinate system changes, while provisioning (see \ref{subsection:provisioning}) provides the tooling to achieve these changes.  Orchestration naturally relies on state management to provide desires as well as current state information.  Hence, orchestration logically sits between state management and provisioning.  In any practical HPC systems management solution, these components may exist as part of one or more tools, but we consider these distinct roles.

\subsubsection{Manageability \& Serviceability}
Orchestration should provide the mechanism for many management tasks on the system.  For instance, it is expected that orchestration will provide a reliable way to implement security patches and software updates across a system.  In this sense, the orchestration system is critical to managing the HPC system.  The orchestration system should also work with the provisioning system to provide reliable, timely system startup and shutdown, including validation of system readiness state and automated recovery from recoverable divergencies in the orchestration task, such as failures in node initialization.

\subsubsection{Scalability \& Resiliency}

As the coordinator of the system, orchestration is a critical service, and failure of the orchestration system will constitute a system management failure.  To build a resilient HPC management solution, the orchestration service must be designed to resist failure as well as provide well-defined behaviors in case of failure and a streamlined recovery process.  To resist failure, orchestration should be capable of replication of service.  This could take the form of tree-like or sibling replication.  It should also be the case that, in the event of an orchestration failure, running work on the system is not interrupted.  Specifically, a failure in the orchestration system should not lead to a failure in a running job or a failure in other system-critical services.  Finally, in the event of a failure, the orchestration system must contain the logic to either resume or, if safe, repeat the current orchestration tasks on the system.

Depending on the complexity of orchestration and the size and complexity of the system, it may not be sufficient to have a single orchestration service managing all nodes in the system.  In this case, it is again desirable to provide either tree-like or active-active sibling orchestration services. This will need to be coupled with tree-like or sibling replication of state management.  Depending on state consistency requirements, mechanisms for replication are well studied \cite{lamport2019part,van2004chain,arnold2006tree}.

\subsubsection{Modern methods}

The last decade has seen many advances in orchestration and automation methods.  Some lessons can be learned from container and virtual machine orchestration solutions, such as Docker \cite{docker}, Kubernetes \cite{kubernetes}, and OpenStack\cite{openstack}.  Orchestration solutions are much more powerful and flexible if they can provide modular and extensible interfaces.  This is greatly assisted by providing well-defined and portable application programming interfaces (APIs), such as OpenAPI/Swagger\cite{swagger}, gRPC\cite{grpc}, or Thrift\cite{thrift}. 

\subsubsection{Implementation}
Orchestration can take many different forms, and can be of varying levels of complexity.  The orchestration system may provide full automation of the system and its allocation, or a minimal subset of coordination tasks.  Some basic features should be required for HPC system management.

\begin{description}
  \item [System initialization \& shutdown] The orchestration system should work with state management and provisioning to provide robust, reliable, and timely system startup and shutdown.  Ideally, the orchestration system should be capable of ordered startup and shutdown of dependent components.  It should also be able to verify system readiness state for these operations and provide basic reporting of operation success or failure, optionally providing automatic recovery from known failure patterns.
  \item [System updates] The orchestration system is responsible for coordinating system updates.  When possible, the orchestration system should provide the ability to roll updates to the system as components become available for updating.
  \item [Security \& emergency remediation] The orchestration system must provide coordination of emergency patching and emergency event remediation, such as controlling firewall rules, revoking access, and emergency software patching.
  \item [System-wide automation flows] The system should provide system-wide automation capabilities both for feature changes, such as loading a new container image for a runtime job, mounting or unmounting filesystems, etc., as well as for automated error recovery.  Ideally, these automation flows should be extensible and definable by the systems administrators.
\end{description}

\subsubsection{Current state}

While orchestration and automation systems have been widely leveraged in the cloud and web-scale computing space, they have had only minimal adoption in HPC.  Container and virtual machine orchestration platforms, such as Kubernetes have been leveraged for certain components of system management in some modern solutions such as Cray Shasta \cite{shasta}, but over-all HPC system orchestration is currently lacking.  Some newer projects, such as Kraken \cite{kraken} have set out to provide more robust orchestration and automation to HPC, but have yet to see wide-spread adoption.  Most HPC system boot models, for instance, still use a ``set it and leave it'' approach to system initialization, in which the provisioning system is put in place, and the process of bringing the system up is left to manually powering on components and verifying that they properly initialize by hand.  We see system orchestration as one of the areas in critical need of future work.

\subsection{Provisioning}
\label{subsection:provisioning}

Provisioning encompasses the processes and tools to discover physical hardware, generate OS images and configuration, boot all node types, and transfer the image and configuration to the nodes.

\subsubsection{Manageability \& Serviceability}

Provisioning is a force multiplier for HPC system administration staff. It allows the management of a single node to effectively require the same amount of time and effort as 1000s of nodes.

Provisioning tooling includes the functionality to automate the configuration of node firmware (e.g.~BIOS settings, BMC settings, etc.), interaction with a network-based boot loader to configure and provide the kernel and initial root filesystem (e.g.~initrd), and provides a mechanism to transfer container and other filesystem images. The generation and configuration of images are also needed.

Reliable, common protocols like HTTPS and SSH should be used as a foundation to build provisioning tooling, as they are well known and easier to troubleshoot. Protocols like HTTPS and IPv6 SLAAC\cite{RFC4862}, which don't require centralized state coordination, are also inherently easier to understand and troubleshoot.

To further reduce complexity, when booting a Linux kernel fetched over the network, using the same kernel through the life-cycle of the boot until the node is rebooted will reduce the overall code base used. Compared to boot processes that kexec from one kernel to another, there is no need to maintain a provisioning kernel that is required to support all hardware within a system.

As described in Section \ref{subsection:minimal-os}, moving from a staged provisioning approach (ie. initrd that makes use kexec or switch\_root to execute the final kernel or root filesystem) to a container layered approach reduces the overall size of boot images and configuration required during boot. Reducing the amount of configuration required during boot inherently makes for an easier to understand and troubleshoot boot process. Avoiding an initrd stage, which with the staged approach isn't accessible after boot, removes a traditionally difficult to troubleshoot step of the boot.

\subsubsection{Scalibility \& Resiliency}

Fast system boot time is often a priority for HPC centers. While a significant part of the boot time of a node lies in executing the system's initial firmware, the network boot loader, network addressing, image transfer, and boot of the image needs to be optimized for speed as well. Reducing boot time increases the overall utilization of the HPC resource, as it allows the node to be more available for jobs. The fast boot of nodes will allow for reboots to be used as part of job cleanup, reducing the complexity and increasing the thoroughness. Administration staff benefit from a fast boot time as well, decreasing the iteration time required when troubleshooting images and other configuration changes.

Node discovery and network addressing should be derived by some inherent facts about a node's physical location within the system (e.g.~based on a switch port and switch ID). A node in a specific location within a given system, should always have the same hostname and network addressing. This approach allows for ease of location finding during maintenance activities as well as a reduction of centrally managed state of a node. For example, a discovery mechanism that records a node's unique MAC address on initial install and subsequent replacement is no longer required when this type of self-discovery functionality exists.

The services that support node boot, addressing, and image transfer need to be horizontally scalable (e.g.~using a load balancer) to support an arbitrarily large number of nodes. Conversely, these services also need to be able to scale down to a single service node or virtual machine for testbeds and other small clusters.

In general, when network-based services and protocols are designed or selected, ones that minimize the amount of communication are preferred. While such ``chatty'' protocols may be acceptable at a smaller scale, when multiplied by 1000 or 10000 nodes, extra consideration is needed to ensure effective scaling.

The use of stateless protocols is preferred for scale-up due to inherit scalability properties as they tend to remove centralized or distributed coordination. An example of a stateless service is IPv6 Stateless Address Autoconfiguration (SLAAC), where the client self-assigns its address based on router advertisements. The router doesn't maintain what address was assigned to the node. An example of a stateful service is DHCP, where the client requests information from the server, the server then maintains state (e.g.~DHCP lease) about the client so when asked again it can give the same response.

\subsubsection{Security}

Provisioning strategies typically rely on hosting artifacts, where artifacts are both configurations (dynamic or statically generated), binaries like the Linux kernel, as well as filesystem images. This model makes it difficult to securely host any type of credential (e.g.~TLS key) that an eventual user on the system won't also have access to over the network. Various strategies have been applied, like a dedicated provisioning VLAN, which isn't configured by the time the node is available for users. Another strategy could be to store credentials in a Trusted Platform Module (TPM), which attempts to make the extraction of a credential from a node difficult.

Most strategies, however, rely on a credential that somehow has to be transferred, prompted for during boot, or stored in non-volatile storage on a node. As a result, the model of out-of-band provisioning is suggested. In this model, an out-of-band processor, commonly known as a baseboard management controller (BMC), provides shared storage media between the node and it. Alternatively, if the BMC has DMA access to the node, it may be possible to create virtual storage devices or simply write artifacts into memory during boot. The advantage of using the BMC, with its dedicated management network, is it can provide a one-way interaction with the node. The user on the node, even with elevated permissions, will be unable to fetch remote artifacts containing credentials. This approach could have the secondary benefit of staging artifacts out of band from the node. These artifacts once staged can then be used to quickly update to a new OS image or kernel for example.

Beyond credentials, BIOS and BMC firmware and settings also need to be protected from elevated privileged users, and thus should only be writable via the out-of-band BMC. In-band mechanisms need to be limited to read-only access or no access at all.

Further, each layer of the provisioning process within a node should cryptographically validate that the layer beneath matches what is expected to be running. Based on the values discovered the node needs to be able to remotely attest\cite{remoteattest} this validation to a centralized service. This and other techniques discussed in Section \ref{subsection:configuration-management} help with tracking change in an HPC system, reducing time to troubleshoot both for administration staff and end-users; as well as allowing sites to be reasonably confident that nodes are running unmodified versions of firmware, OS, and software.

\subsubsection{Implementation}

\begin{description}

\item[Node Discovery] Network addressing strategies should employ self-discovery based on the physical location in a system. For example, an addressing scheme can be developed based on the connected switch port ID and switch chassis ID fetched via Link Layer Discovery Protocol (LLDP). Another approach is using a DHCP Relay running on the node connected switch, where the relay includes circuit ID and port ID information in the forwarded DHCP request.

With IPv6, a network can be configured so that a node solicits and receives IPv6 router advertisements including RDNSS (DNS Servers) and DNSSL (DNS Search Path) settings for initial bootstrapped addressing. The resulting address is typically derived from the hardware address of the network adapter. Based on the advertisement the node will have an IP address and DNS server and search path information. Further configuration attributes for boot can be discovered via DNS. The advantage of this approach over DHCP is there is no centrally maintained state during the network address discovery process.

Note, while these network addressing strategies are important for scaling to large systems, it's also important to support the basic case for a small cluster using traditional DHCP and PXE without regard to a physical location. This keeps the required network configuration to a minimum for testbeds and smaller systems.

\item[Image Building] There is a range of image building options available. It is not clear which approach below is the best, and likely the pros and cons of each will be dependent on use cases. Thus it's suggested that these options be studied to better describe the benefits and deficiency of the following imaging approaches.

\begin{enumerate}
  \item At one extreme, the fewest number of images are created, a generalized image is created ahead of the node boot. This image is modified on boot with configuration management, or another runtime tooling to customize it for the node's specific purpose.
  \item At the other end of the spectrum, an individual image is created for every single node; images are completely customized before the boot including all needed configuration.
\end{enumerate}

Other ideas within the above extremes should be studied. For example, perhaps node-specific configuration isn't part of the image, but is still generated ahead of time. This configuration is then applied to a node via an overlay mechanism during boot.

\item[OS Image Transfer] Based on the system design, administrator reasoning, or user choice two image transfer mechanisms are needed.

\begin{enumerate}
  \item Transfer of the entire image into node memory or local storage. The main drawback is occupying memory and thus reducing the total amount of memory available to jobs. The overall data transferred over the network is maximized as the entire content of the image is always transferred.
  \item Partially transfer the image into node memory or local storage, with the remaining contents of the image remotely available. Remote contents of the image are fetched (and potentially cached) when explicitly read by the node. This can be implemented with existing tools like block-based storage (e.g.~iSCSI and Ceph RBD), or network-based filesystems (e.g.~NFS or Lustre). Each strategy typically mounts the remote storage read-only with an overlay filesystem that allows local changes to be applied (but not persisted).
\end{enumerate}

\end{description}

\subsubsection{Current State}

The primary HPC-focused provisioning tools currently are xCat\cite{xcat}, Warewulf{\cite{warewulf}}, vendor-provided tools such as those provided by HPE/Cray\cite{shasta}, as well as numerous cloud and enterprise-focused tools like Cobbler\cite{cobbler}, Foreman\cite{foreman}, MAAS (Metal as a Service)\cite{MAAS}, Digital Rebar\cite{digitalrebar}, and OpenStack Ironic\cite{openshift}. These tools all implement some form of configuration automation for DHCP and the network boot loader iPXE or PXELINUX. The HPC tools generally focus on image-based provisioning while the enterprise-focused tools typically focus on Linux distribution specific install and configuration like Kickstart\cite{kickstart} and Preseed\cite{debianhandbook}. While these tools may implement parts of the concepts presented in Section \ref{section:node-stack}, we are unaware of an existing integrated provisioning tool or tools that implement all concepts.

\section{Conclusions}

The HPC system software stack has relied on incremental changes over the last decade to reach the scales we currently see in HPC systems.  However, by taking a critical look at the current state of the art within our field and parallels within web companies, we can see a variety of practices that can be beneficial to adapt to make HPC systems more manageable, serviceable, scalable, resilient, and secure.

In Section \ref{section:node-stack}, we looked at how using a minimal operating system with container-based services and jobs can lead to less downtime, greater flexibility, and greater scalability.  While few HPC-specific services have been designed with this model in mind, we believe that many can translate to this model with minimal effort due to their horizontal scaling features.  Methods of containerizing jobs, on the other hand, are in active development, but are still relatively immature when compared to their traditional counterparts.  We see a lot of potential in moving toward containerized workflows in both of these areas.

In Section \ref{section:system-management}, we took a more abstract look at the system software stack.  We discussed concepts that will help modernize configuration management, state management, orchestration, and provisioning within an HPC system, making recommendations for how to make better use of existing tools and adopt modern practices with new tools as they are developed.  These are areas that can benefit from new and continued development.

The HPC system software stack is huge, and a paper of this length can only scratch the surface of potential work.  Continued work in this area within the HPC community as a whole is required to realize the gains described in this paper and the many similar gains that are waiting to be described.

\begin{acks}
This research used resources of the Argonne Leadership Computing Facility at Argonne National Laboratory, which is supported by the Office of Science of the U.S. Department of Energy, Office of Science, under contract number DE-AC02-06CH11357.

This work was supported by the U.S. Department of Energy through the Los Alamos National Laboratory. Los Alamos National Laboratory is operated by Triad National Security, LLC, for the National Nuclear Security Administration of U.S. Department of Energy (Contract No. 89233218CNA000001).

This research used resources of the National Energy Research Scientific Computing Center (NERSC), a U.S. Department of Energy Office of Science User Facility operated under Contract No. DE-AC02-05CH11231.

This research used resources of the Oak Ridge Leadership Computing Facility at the Oak Ridge National Laboratory, which is supported by the Office of Science of the U.S. Department of Energy under Contract No. DE-AC05-00OR22725.
\end{acks}

\bibliographystyle{ACM-Reference-Format}
\bibliography{03-bibliography}

\end{document}